\documentclass[preprint,showpacs,preprintnumbers,amsmath,amssymb,floatfix]{revtex4}

\usepackage{graphicx}
\usepackage{dcolumn}
\usepackage{bm}

\begin{document}


\title{Investigation of excited states in $^{18}$Ne via resonant elastic scattering of $^{17}$F+p and its astrophysical implication in
the stellar reaction of $^{14}$O($\alpha$,$p$)$^{17}$F}

\author{J. Hu$^{1,2}$}
\author{J.J.~He$^1$}
\altaffiliation[Correspondence Author: ]{jianjunhe@impcas.ac.cn}
\author{S.W. Xu$^1$}
\author{Z.Q. Chen$^1$}
\author{X.Y. Zhang$^1$}
\author{J.S. Wang$^1$}
\author{X.Q. Yu$^1$}
\author{L. Li$^{1,2}$}
\author{L.Y. Zhang$^{1,2}$}
\author{Y.Y. Yang$^{1,2}$}
\author{P. Ma$^{1,2}$}
\author{X.H. Zhang$^1$}
\author{Z.G. Hu$^1$}
\author{Z.Y. Guo$^1$}
\author{X. Xu$^{1,2}$}
\author{X.H. Yuan$^{1}$}
\author{W. Lu$^{1,2}$}
\author{Y.H. Yu$^{1}$}
\author{Y.D. Zang$^{1,2}$}
\author{S.W. Tang$^{1,2}$}
\author{R.P. Ye$^{1,2}$}
\author{J.D. Chen$^{1,2}$}
\author{S.L. Jin$^{1,2}$}
\author{C.M. Du$^{1,2}$}
\author{S.T. Wang$^{1,2}$}
\author{J.B. Ma$^{1,2}$}
\author{L.X. Liu$^{1,2}$}
\author{Z. Bai$^{1,2}$}
\author{X.G. Lei$^1$}
\author{Z.Y. Sun$^1$}
\author{Y.H. Zhang$^1$}
\author{X.H. Zhou$^1$}
\author{H.S. Xu$^1$}

\affiliation{
$^{1}$Institute of Modern Physics (IMP), Chinese Academy of Sciences, Lanzhou 730000, China\\
$^{2}$Graduate School of Chinese Academy of Sciences, Beijing 100049, China}

\author{J. Su$^3$}
\author{E.T. Li$^3$}
\affiliation{$^{3}$China Institute of Atomic Energy (CIAE), P.O. Box 275(46), Beijing 102413, China}

\author{H.W. Wang$^4$}
\author{W.D. Tian$^4$}
\affiliation{$^{4}$Shanghai Institute of Applied Physics (SINAP), Chinese Academy of Sciences, Shanghai 201800, China}

\author{X.Q. Li$^5$}
\affiliation{$^{5}$School of Physics and State Key Laboratory of Nuclear Physics and Technology, Peking University, Beijing 100871, China}

\date{\today}  

\begin{abstract}
Properties of proton resonances in $^{18}$Ne have been investigated efficiently by utilizing a technique of proton resonant elastic scattering with a $^{17}$F
radioactive ion (RI) beam and a thick proton target. A 4.22~MeV/nucleon $^{17}$F RI beam was produced via a projectile-fragmentation reaction, and subsequently
separated by a Radioactive Ion Beam Line in Lanzhou ({\tt RIBLL}).
Energy spectra of the recoiled protons were measured by two sets of $\Delta$E-E silicon telescope at center-of-mass scattering angles of
$\theta_{c.m.}$$\approx$175${^\circ}$$\pm$5${^\circ}$, $\theta_{c.m.}$$\approx$152${^\circ}$$\pm$8${^\circ}$, respectively.
Several proton resonances in $^{18}$Ne were observed, and their resonant parameters have been determined by an $R$-matrix analysis of the differential cross sections
in combination with the previous results.
The resonant parameters are related to the reaction-rate calculation of the stellar $^{14}$O($\alpha$,$p$)$^{17}$F reaction, which was thought to be the breakout
reaction from the hot CNO cycles into the $rp$-process in x-ray bursters.
Here, $J^\pi$=(3$^-$, 2$^-$) are tentatively assigned to the 6.15-MeV state which was thought the key 1$^-$ state previously.
In addition, a doublet structure at 7.05 MeV are tentatively identified, and its contribution to the resonant reaction rate of $^{14}$O($\alpha$,$p$)$^{17}$F could be
enhanced by at least factors of about 4$\sim$6 in comparison with the previous estimation involving only a singlet. The present calculated resonant rates are much
larger than those previous values, and it may imply that this breakout reaction could play a crucial role under x-ray bursters conditions.
\end{abstract}



\maketitle

\section{Introduction}
Explosive hydrogen and helium burning are thought to be the main sources for energy generation and nucleosynthesis of heavier elements in cataclysmic
binary systems~\cite{bib:woo76,bib:cha92,bib:wie98}.
In such a close binary system, hydrogen and helium rich material from a companion star pile up onto the surface of a neutron star, and form an accretion disk where
thermal runaway reactions can be ignited through both the triple-$\alpha$ reaction and breakout from the hot carbon-nitrogen-oxygen (CNO) cycles into the rapid proton
capture process ($rp$-process). Energy generation increases rapidly as a function of temperature, and hence the rate of energy release can increase faster than the
rate of cooling, ultimately leading to uncontrollable thermonuclear explosions in the accreting disk, the so called x-ray bursts ({\em e.g.}, Type I x-ray
bursts~\cite{bib:taa85}).
The $\alpha$$p$ chain is initiated through the reaction sequence $^{14}$O($\alpha$,$p$)$^{17}$F($p$,$\gamma$)$^{18}$Ne($\alpha$,$p$)$^{21}$Na~\cite{bib:bar00}, and
increases the rate of energy generation by 2 orders of magnitude~\cite{bib:wie98}. In x-ray burster scenarios, the nucleus $^{14}$O ($t_{1/2}$=71 s) forms an important
waiting point, and the ignition of the $^{14}$O($\alpha$,$p$)$^{17}$F reaction at temperatures $\sim$0.4 GK produces a rapid increase in power and can lead to breakout
from the hot CNO cycles into the $rp$-process with the production of medium mass proton-rich nuclei~\cite{bib:sch98,bib:sch01,bib:bre09}.
As a crucial breakout reaction, its reaction rate determines the conditions under which the bursts are initiated and triggered, and thus plays an important role
in the field of nuclear astrophysics.

Wiescher {\em et al.}~\cite{bib:wie87} calculated the reaction rates of $^{14}$O($\alpha$,p)$^{17}$F in detail, and showed that the resonant reaction rates
dominated the total rates above temperature $\sim$0.4 GK. Later on, Funck {\em et al.}~\cite{bib:fun88,bib:fun89} found that direct-reaction contribution to the $\ell$=1
partial wave was comparable to or even greater than the resonant one at certain temperatures. Since resonant reaction rates of $^{14}$O($\alpha$,p)$^{17}$F
depend sensitively on the resonant energies, spin-parities, partial and total widths of the relevant excited states in the compound nucleus $^{18}$Ne,
Hahn {\em et al.}~\cite{bib:hah96} extensively studied the levels in $^{18}$Ne by three reactions, $^{16}$O($^3$He,$n$)$^{18}$Ne, $^{12}$C($^{12}$C,$^6$He)$^{18}$Ne
and $^{20}$Ne($p$,$t$)$^{18}$Ne. It was found that this reaction rate, at temperature above $\sim$0.5 GK, was dominated by capture on a single 1$^-$ resonance at an
excitation energy of 6.150 MeV lying 1.035 MeV above the $^{14}$O+$\alpha$
threshold ($Q_\alpha$=5.115 MeV~\cite{bib:aud03}). Harss {\em et al.}~\cite{bib:har99} studied the time reverse reaction $^{17}$F($p$,$\alpha$)$^{14}$O by using a
$^{17}$F beam at {\tt ANL}, and determined the resonance strengths for three levels at 7.16, 7.37, 7.60 MeV.
Later, G\'{o}mez del Campo {\em et al.}~\cite{bib:gom01} used a $p$($^{17}$F,$p$) resonant elastic scattering on a thick CH$_2$ target to look for resonances of
astrophysical interest in $^{18}$Ne at {\tt ORNL}, and assigned the $E_x$=6.15, 6.35 MeV states as J$^{\pi}$=1$^-$, and 2$^-$, respectively.
Subsequently, a new set of resonant parameters [$E_r$, $J^\pi$, $\Gamma$, {\em etc.}] for several resonances in $^{18}$Ne were deduced from another experiment at
{\tt ANL}~\cite{bib:har02}, and the spin-parities were reassigned based on the Coulomb-shift calculation as well as Fortune and Sherr's comments~\cite{bib:for00}.
The resonance strength and $\Gamma_\alpha$ width for the 6.15-MeV state was extracted based on a 1$^-$ assignment as well as the previously obtained excitation
function~\cite{bib:har99}.
In addition, the inelastic component of this key 1$^-$ resonance in the $^{14}$O($\alpha$,p)$^{17}$F reaction was also studied by a new highly sensitive technique at
{\tt ISOLDE/CERN}~\cite{bib:hjj09,bib:hjj10}. It was found that this inelastic component would enhance the reaction rate, contributing approximating equally to the
ground-state component of the reaction rate, however not to the relative degree suggested in Ref.~\cite{bib:bla03}.

Although our understanding in the reaction rates of $^{14}$O($\alpha$,p)$^{17}$F has been greatly improved so far, there are still some discrepancies. Recently, the
1$^-$ assignment for the 6.15-MeV state was questioned~\cite{bib:he10} by a careful reanalysis of the previous experimental data~\cite{bib:gom01}. It has been found
that most probably the 6.286-MeV state is the key 1$^-$ state whereas the 6.15-MeV state is a 3$^-$ or 2$^-$ state, and hence the resonance at $E_x$=6.286 MeV probably dominates
the reaction rates in the temperature below $\sim$2 GK~\cite{bib:he10}. However, low statistics of experimental data~\cite{bib:gom01} around $E_x$=6.286 MeV prevent us
from putting the new assignments on a very firm ground. In addition, a thick target measurement was performed in a direct study of the $^{14}$O($\alpha$,p)$^{17}$F
reaction~\cite{bib:not04,bib:not041,bib:kub06}, where a new peak near $E_{c.m.}^\alpha$$\sim$1.5 MeV was observed and was
thought to be an inelastic branch of $^{14}$O($\alpha$,p)$^{17}$F reaction proceeding through a resonance at $E_x\sim$7.1 MeV in $^{18}$Ne which decays
to the first excited state in $^{17}$F ($E_x$=0.495 MeV) by proton emission. According to the estimation, this inelastic branch would increase the astrophysical
$^{14}$O($\alpha$,p)$^{17}$F rate by $\sim$50\% at temperatures beyond 2 GK. New {\tt ORNL} experimental result~\cite{bib:bar10} however, has ruled out this
inelastic-branch interpretation.

In the present work, we have investigated the proton resonant properties of the compound nucleus $^{18}$Ne, covering an excitation energy ($E_x$) region of
4.7$\sim$8.0 MeV, by using a resonant elastic scattering of a $^{17}$F beam with a thick (CH$_2$)$_n$ target. This sort of thick-target
method~\cite{bib:art90,bib:kub01,bib:ter07,bib:he07,bib:he09} enables simultaneous measurements of excitation function within a very wide energy range.
The proton resonant parameters have been determined by an $R$-matrix analysis of the differential cross sections of $^{17}$F+$p$. The astrophysical implication of
$^{14}$O($\alpha$, $p$)$^{17}$F reaction is briefly discussed based on the present work.

\section{Experiment}
The experiment was carried out by using the Heavy Ion Research Facility in Lanzhou ({\tt HIRFL})~\cite{bib:xia02,bib:zha10}.
A primary beam of $^{20}$Ne$^{10+}$ was accelerated up to 69.5 MeV/nucleon by a Separate Sector Cyclotron ({\tt SSC}, $K$ = 450) with an intensity of $\sim$300 enA,
and bombarded a 4094-$\mu$m $^9$Be primary target. A secondary beam of $^{17}$F$^{9+}$, which was produced via a projectile-fragmentation reaction mechanism,
was then separated, purified, and transported by a Radioactive Ion Beam Line in Lanzhou ({\tt RIBLL})\cite{bib:sun03} to the secondary target chamber.

\begin{figure}
\begin{center}
\includegraphics[width=8cm]{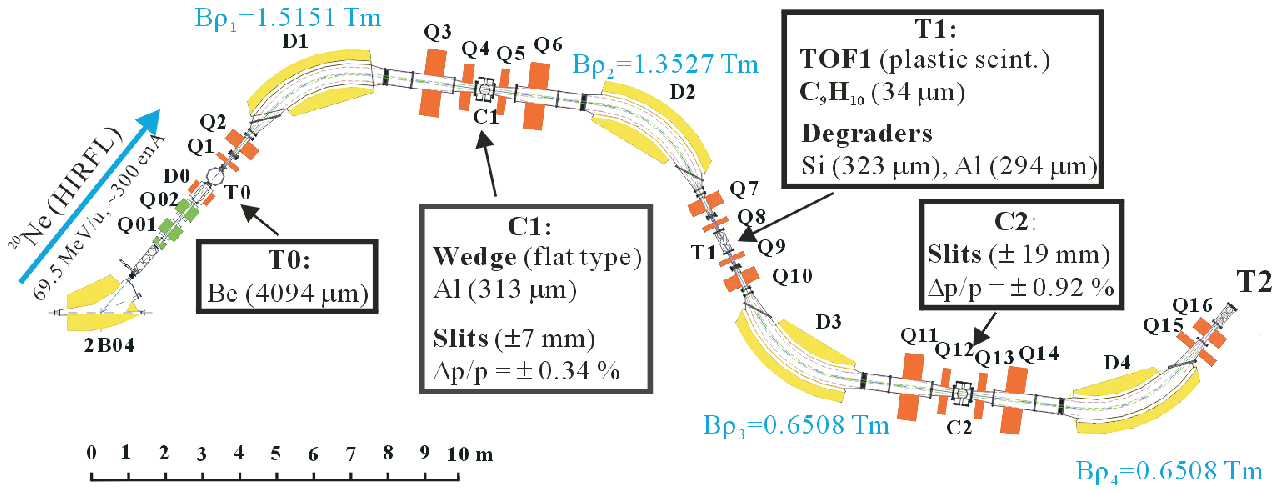}
\end{center}
\caption{\label{fig1} Schematic view of the $\texttt{RIBLL}$. All relevant parameters are indicated in the figure.}
\end{figure}

The schematic view of {\tt RIBLL} is shown in Fig.~\ref{fig1}.
At the momentum-dispersive focal plane ($\textbf{C1}$), a 313-$\mu$m flat-shaped Al degrader was installed to separate the $^{17}$F$^{9+}$ particles from other
reaction products. In order to meet physics requirement, the particle energies were subsequently reduced by two degraders (a 323-$\mu$m Si and a 294-$\mu$m Al)
and one plastic scintillator (a 34-$\mu$m C$_9$H$_{10}$ foil, hereafter referred to as {\tt TOF1}) at the intermediate focal point ($\textbf{T1}$). Two horizontal slits
were used to purity the beam and to restrict the momentum spreads of the $^{17}$F$^{9+}$ particles, here, the momentum spread $\Delta p/p$ was limited to $\pm$0.34\% at
$\textbf{C1}$ and to $\pm$0.92\% at $\textbf{C2}$, respectively.

\begin{figure}
\begin{center}
\includegraphics[scale=0.5]{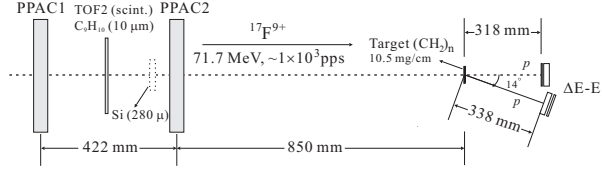}
\end{center}
\caption{\label{fig2} Experimental setup for the scattering measurement at $\texttt{T2}$. See text for details.}
\end{figure}

\begin{figure}
\begin{center}
\includegraphics[scale=0.5]{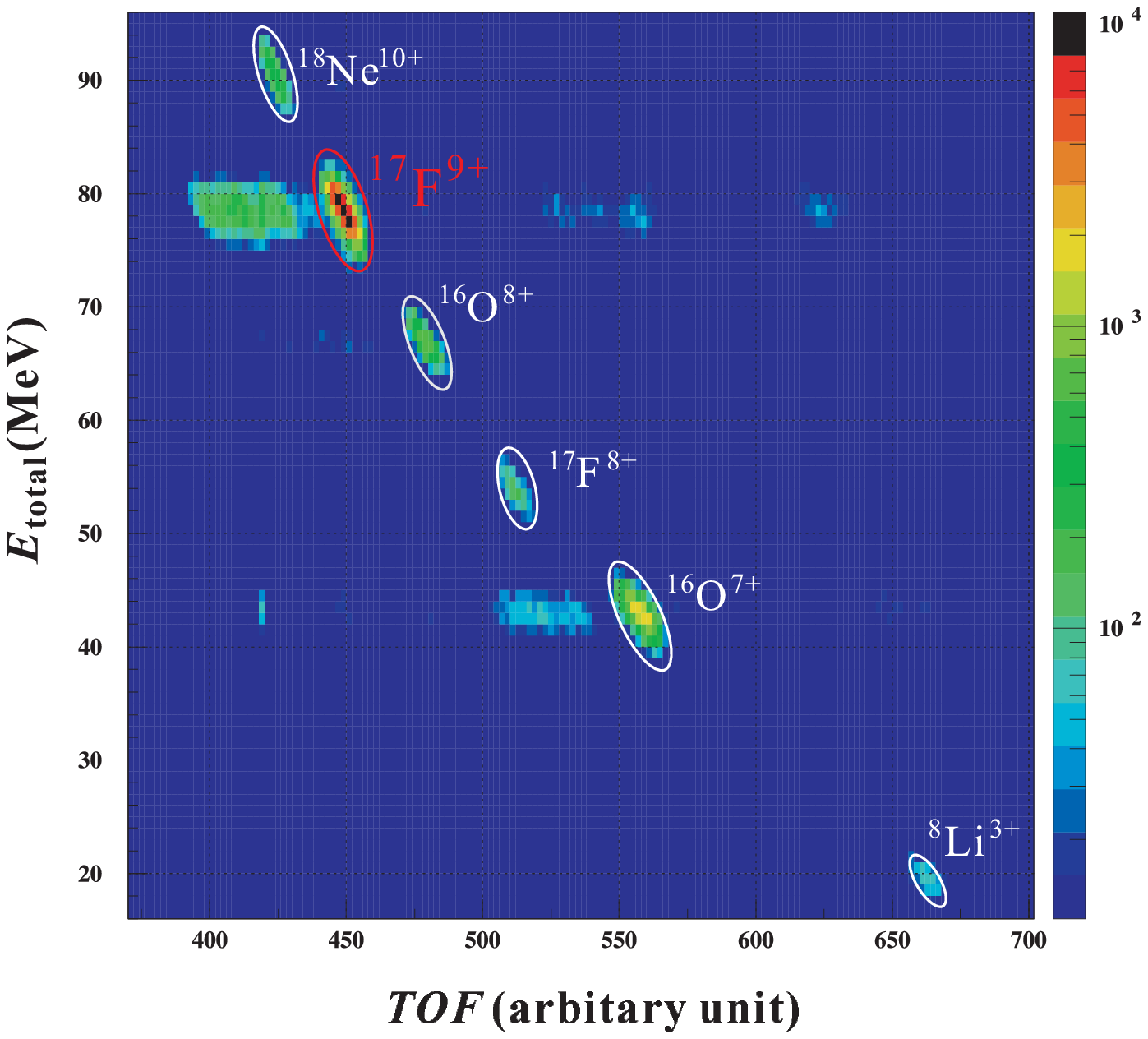}
\end{center}
\caption{\label{fig3} Identification plot for the beam particles. Where, $E_\mathrm{total}$ indicates the total beam energy deposited in the 280-$\mu$m Si detector at
$\texttt{T2}$ (just before $\texttt{PPAC2}$), and $TOF$ indicates the time-of-flight measured by $\texttt{TOF1}$ \& $\texttt{TOF2}$ detectors at $\textbf{T1}$ and
$\textbf{T2}$. See text for details.}
\end{figure}

\begin{figure}
\begin{center}
\includegraphics[scale=0.6]{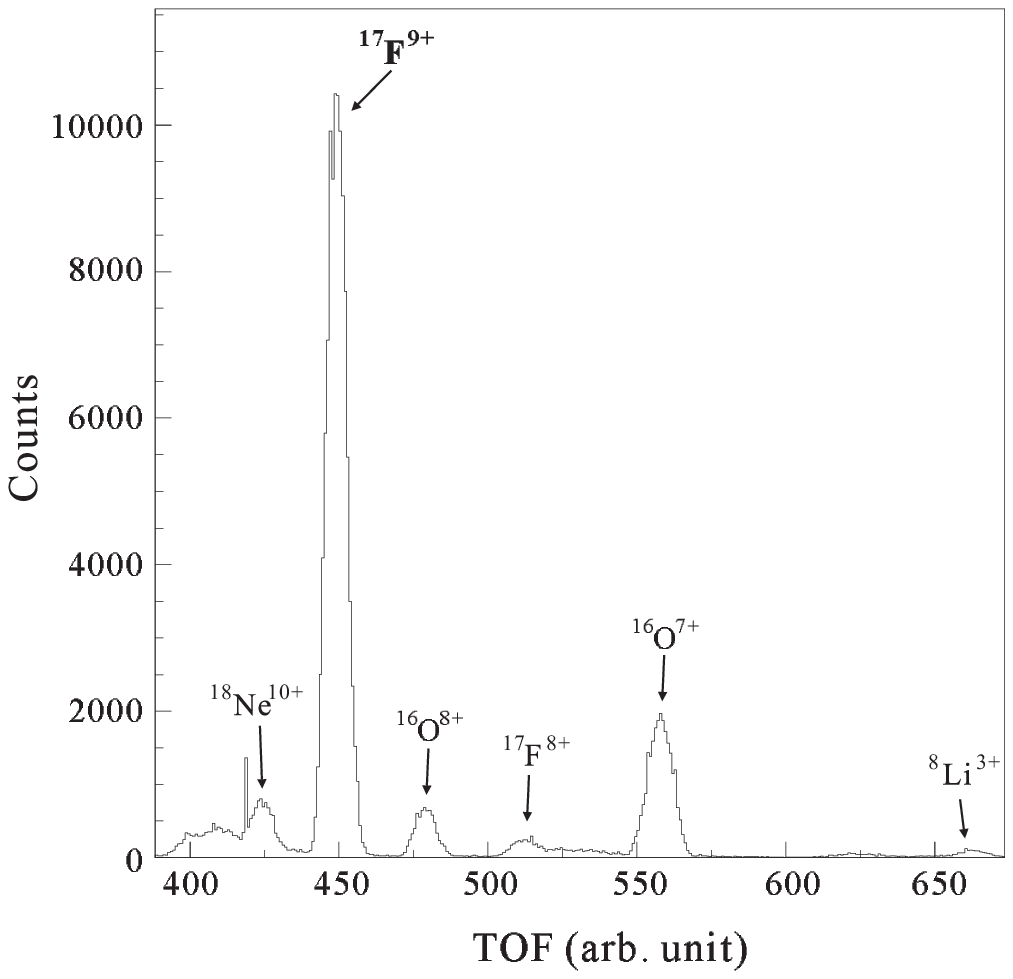}
\end{center}
\caption{\label{fig4} Identification plot for the beam particles in a typical TOF spectrum.}
\end{figure}

At the achromatic focal plane ($\textbf{T2}$), a scattering setup was installed inside a vacuum chamber as shown in Fig.~\ref{fig2}. The setup consisted of
two parallel-plate avalanche counters ($\texttt{PPACs}$)\cite{bib:ma10}, one plastic scintillator (a 10-$\mu$m C$_9$H$_{10}$ foil, hereafter referred to as {\tt TOF2}),
one 10.5-mg/cm$^{2}$ (CH$_{2})_{n}$ target and two sets of $\Delta$E-E silicon telescope detection system. During the beam tuning, a 280-$\mu$m Si detector was
inserted just before $\texttt{PPAC2}$ to measure the total energies of the particles (see the dashed box in Fig.~\ref{fig2}). All beam particles can be identified
clearly by using the time-of-flight (TOF) measured by the $\texttt{TOF1}$ and $\texttt{TOF2}$ scintillators, together with the total energy deposited in the Si detector
in an event-by-event mode. Figure~\ref{fig3} shows an identification plot for the beam particles, and it can be seen that the $^{17}$F$^{9+}$ particles can be
identified completely by using the TOF information only. Although this Si detector was moved out during the experimental runs, the $^{17}$F$^{9+}$ particles can be
identified uniquely by using the TOF information as shown in Fig.~\ref{fig4}. The position and incident angle of the beam particles on the target have been determined
by extrapolating the two-dimensional hit positions measured by two $\texttt{PPAC}$s whose position resolution is about 1 mm (FWHM)~\cite{bib:ma10}.

At (CH$_{2})_{n}$ target position, the average intensity of $^{17}$F$^{9+}$ beam was about 1$\times$10$^{3}$ particles/s with a $\sim$50\% purity.
The $^{17}$F$^{9+}$ beam-spot widths (FWHM) were 17 mm horizontally and 16 mm vertically. The horizontal and vertical angular spread (FWHM) of the beam were 30 mrad
and 9 mrad, respectively. The mean energy of $^{17}$F$^{9+}$ was 4.22 MeV/nucleon with a 0.25 MeV/nucleon width (FWHM).
In addition, the experimental data with a C target (10.7 mg/cm$^{2}$) was also acquired in a separate run to evaluate the contributions from the reactions of $^{17}$F
with C nuclei contained in the (CH$_{2})_{n}$ target.

\begin{figure}
\includegraphics[scale=0.5]{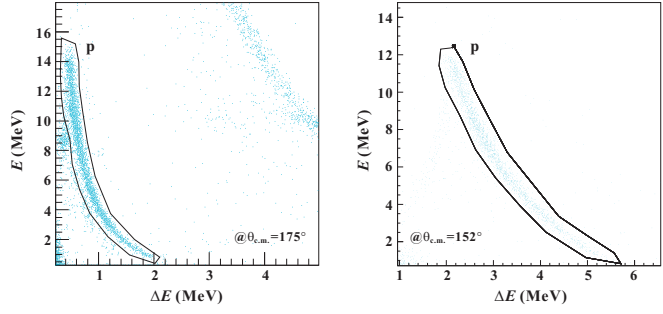}
\caption{\label{fig5} Identification plot for the recoiled particles in two $\Delta$E-E telescopes. The polygonal boxes for the proton gates are shown and
utilized in the analysis.}
\end{figure}

All beam particles were fully stopped in the thick (CH$_{2})_{n}$ target.
The light particles recoiled from the target were measured by using two sets of $\Delta$E-E Micron silicon telescope~\cite{bib:micron} at scattering angles,
$\theta_{lab}$$\approx$0${^\circ}$, 14${^\circ}$, respectively (see Fig.~\ref{fig2}). At $\theta_{lab}$$\approx$0${^\circ}$, the telescope was consisted of a
63-$\mu$m-thick W1-type double-sided-strip (16$\times$16 strips) detector and a 1500-$\mu$m-thick MSX25-type pad detector; at $\theta_{lab}$$\approx$14${^\circ}$,
the telescope was consisted of a 300-$\mu$m-thick MSQ25-type detector and a 1000-$\mu$m-thick MSPX042-type double-sided-strip (16$\times$16 strips) detector. Each
silicon detector has a sensitive area of 50$\times$50 mm$^2$ and subtends an angular range of $\Delta \theta_{lab}$$\approx$9${^\circ}$. Two telescopes covered
laboratory solid angles ($\Delta\Omega_{lab}$) about 25 msr and 22 msr, respectively. In the center-of-mass (c.m.) frame, the scattering angles covered by two
telescopes are $\theta_{c.m.}$$\approx$175${^\circ}$$\pm$5${^\circ}$ and $\theta_{c.m.}$$\approx$152${^\circ}$$\pm$8${^\circ}$, respectively.
The recoiled protons were clearly identified by the energies deposited both in $\Delta$E and E detectors as shown in Fig.~\ref{fig5}, where the energy calibration
for detectors was performed by using a standard triple $\alpha$ source.

\section{Results}
Differential cross section for the $^{17}$F+$p$ elastic scattering in the c.m. frame has been calculated by the following relationship
({\em e.g.}, see~\cite{bib:kub01,bib:he07})
\begin{eqnarray}
\frac{d\sigma}{d\Omega_{c.m.}}(E_{c.m.},\theta_{c.m.})=\frac{1}{4 \mathrm{cos} \theta_{lab}} \times \frac{N_p}{I_b N_H \Delta\Omega_{lab}}
\label{eq:two},
\end{eqnarray}
where $N_p$ is the number of detected protons, {\em i.e.}, at energy interval of $E_{c.m.} \rightarrow E_{c.m.}+ \Delta E_{c.m.}$; $I_b$ is the total number of
$^{17}$F$^{9+}$ beam particles bombarding on the (CH$_2)_n$ target; $N_H$ is the number of H atoms per unit area per energy bin ($\Delta E_{c.m.}$) in the
target~\cite{bib:zie85}. Here, center-of-mass energy, $E_{c.m.}$, for the $^{17}$F+$p$ elastic scattering has been calculated by ({\em e.g.}, see~\cite{bib:he07})
\begin{eqnarray}
E_{c.m.}=\frac{A_b + A_t}{4 A_b \mathrm{cos}^2 \theta_{lab}} E_p
\label{eq:one},
\end{eqnarray}
where $A_b$ and $A_t$ are the mass numbers of the beam and target nuclei; the proton energy at the reaction point, $E_p$, has been calculated through the total energy
deposited in the detectors by correcting the proton energy loss in the target. Thus, energy of the excited state in $^{18}$Ne can be calculated by $E_{x}$=$E_{r}$+$S_p$,
thereinto the proton separation energy of $^{18}$Ne is $S_p$=3.924 MeV~\cite{bib:aud03}, and the resonant energy $E_{r}$ has been determined by the $R$-matrix analysis
discussed below. The uncertainty of the deduced excitation energy is estimated to be about $\pm$30 keV.

\begin{figure}
\includegraphics[scale=0.5]{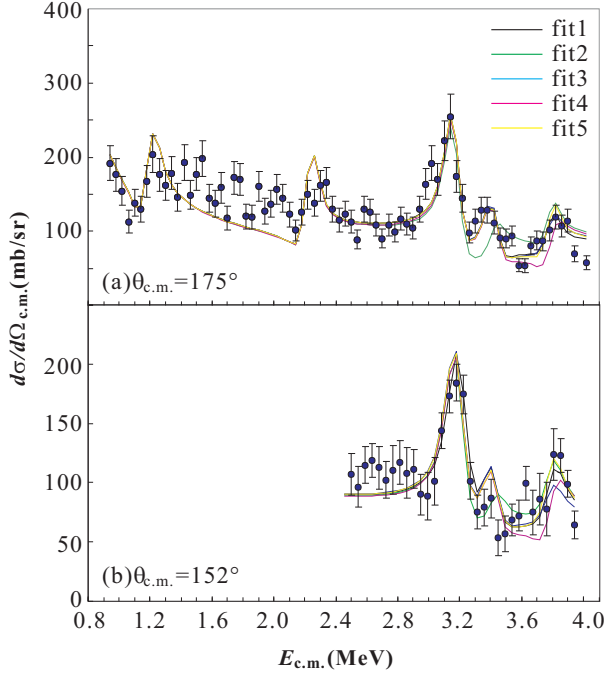}
\caption{\label{fig6} The c.m. differential cross sections for elastically scattered protons produced by bombarding a thick (CH$_2$)$_n$ target with 71.7 MeV $^{17}$F
particles at scattering angles of (a) $\theta_{c.m.}$$\approx$175${^\circ}$$\pm$5${^\circ}$ and (b) $\theta_{c.m.}$$\approx$152${^\circ}$$\pm$8${^\circ}$.
The curved lines represent the $R$-matrix fits to the data, see text for details.}
\end{figure}

Figure~\ref{fig6}(a)\&(b) show the deduced c.m. differential cross sections for the $^{17}$F+$p$ elastic scattering at angles of
$\theta_{c.m.}$$\approx$175${^\circ}$$\pm$5${^\circ}$ and $\theta_{c.m.}$$\approx$152${^\circ}$$\pm$8${^\circ}$, respectively. The uncertainties shown in the figures
are mainly the statistical ones. Note the data shown in Fig.~\ref{fig6}(b) were eliminated at energy region below 2.4 MeV because of the electronics noise.

In order to determine the resonant parameters of the observed resonances, the multichannel $R$-matrix calculations~\cite{bib:lan58,bib:des03,bib:bru02} (see
example~\cite{bib:he10,bib:alex09}) that include the energies, widths, spins, angular momenta, and interference sign for each candidate resonance have been performed
in the present work. A channel radius of $r_0$=1.25 fm [$R$=$r_0$$\times$(1+17$^{1/3}$)] appropriate for the $^{17}$F+$p$ system\cite{bib:wie87,bib:hah96,bib:gom01}
has been utilized in the present $R$-matrix calculation, where the fitting results are insensitive to the choice of radius within the present statistics.
The ground state spin-parity configurations of $^{17}$F and proton are 5/2$^+$ and 1/2$^+$, respectively. Thus, there are two channel spins in the elastic channel,
{\em i.e.} $s$ = 2, 3. Actually there are only minor differences between the $R$-matrix calculations with two different channel spins.

Five resonances, {\em i.e.}, at $E_x$=5.11, 6.15, 7.09, 7.34 and 7.71 MeV, have been analyzed and the fitting curves are shown in Fig.~\ref{fig6}(a)
(at $\theta_{c.m.}$=175$^\circ$). The resonant parameters ($E_x$, $J^\pi$[$\ell$], $\Gamma_p$) used in the $R$-matrix calculations are listed in Table~\ref{table1}.
Here, the parameters are fixed for the first three states in all fits, {\em i.e.},
5.11 MeV ($J^\pi$=2$^+$, $\ell$=0, $\Gamma_p$=45 keV)~\cite{bib:gom01,bib:hah96,bib:hjj09}, 6.15 MeV ($J^\pi$=2$^-$, $\ell$=3, $\Gamma_p$=40 keV)~\cite{bib:he10}, and
7.09 MeV ($J^\pi$=4$^+$, $\ell$=2, $\Gamma_p$=80 keV)~\cite{bib:hah96,bib:har02}. In `fit2' calculation, the fit of a $J^\pi$=1$^-$ ($\ell$=3) assignment to the
7.34-MeV state was not good, and this may possibly exclude the 1$^-$ assignment~\cite{bib:hah96} and support the 2$^+$ assignment~\cite{bib:har02} for the nominal
7.35-MeV state~\cite{bib:hah96}. As for the 7.71-MeV state, $J^\pi$=2$^-$ (`fit1' \&`fit2'), 1$^-$ (`fit3'), 2$^+$ (`fit4'), and 3$^-$ (`fit5') assignments have been
attempted, and all fits are reasonable except the 2$^+$ assignment (`fit4'). The $R$-matrix fitting results at scattering angle $\theta_{c.m.}$$\approx$152$^\circ$
are consistent with those at $\theta_{c.m.}$$\approx$175$^\circ$.

\begin{figure}
\includegraphics[scale=0.8]{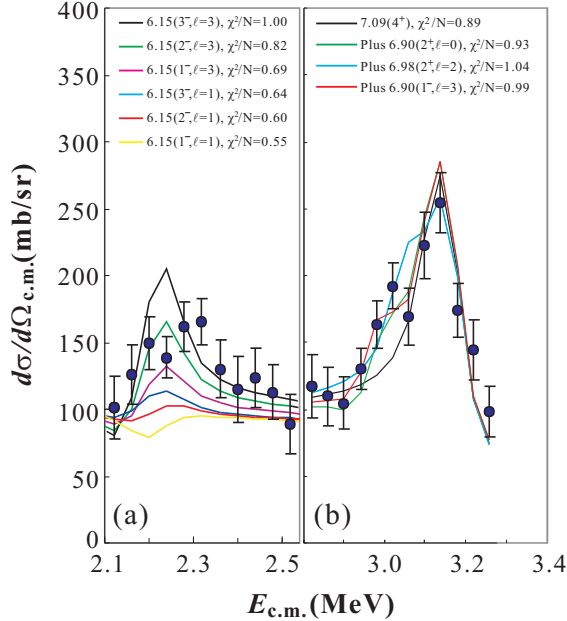}
\caption{\label{fig7} $R$-matrix fit results for the peak at (a) 6.15 MeV, and (b) around 7.09 MeV, respectively. See text for details.}
\end{figure}

Figure~\ref{fig7}(a)\&(b) show detailed $R$-matrix fits to the 6.15- and 7.05-MeV states, respectively.
For the 6.15-MeV state, it shows that $J^\pi$=(3$^-$, 2$^-$) assignments can reproduced the experimental data reasonably with a transferred angular momentum $\ell$=3 but
not with an $\ell$=1~\cite{bib:he10}, and additionally the 1$^-$ assignment (either $\ell$=3 or 1) seems unlikely [see Fig.~\ref{fig7}(a)]. For the 7.05-MeV state, it
was strongly populated by the $^{16}$O($^3$He,$n$)$^{18}$Ne reaction but no by the $^{21}$Ne($p$,$t$)$^{18}$Ne reaction~\cite{bib:ner81,bib:hah96}.
Nero {\em et al.}~\cite{bib:ner81} observed a 7.06-MeV state ($\Gamma$=180$\pm$50 keV) exhibiting $J^\pi$=(1$^-$, 2$^+$) characters in the measured angular distribution,
and thought 2$^+$ was a more admissible assignment based on a reaction-mechanism analysis. Hahn {\em et al.}~\cite{bib:hah96} also observed a broad state at 7.07 MeV
($\Gamma$=200$\pm$40 keV), and achieved a smaller fitting $\chi^2$/N value for a doublet at 7.05\&7.12 MeV. In the present work, the peak around 7.07 MeV exhibits more
like a doublet as seen in Fig.~\ref{fig7}(b) where $R$-matrix fitting results for the doublets are better than that for a singlet.

\begin{table}
\caption{\label{table1} $R$-matrix parameters ($E_x$, $J^\pi$[$\ell$], $\Gamma_p$) used in Fig.~\ref{fig6}. The parameters are fixed for the first three states in all
fits, {\em i.e.}, 5.11 MeV ($J^\pi$=2$^+$, $\ell$=0, $\Gamma_p$=45 keV)~\cite{bib:gom01,bib:hah96,bib:hjj09},
6.15 MeV ($J^\pi$=2$^-$, $\ell$=3, $\Gamma_p$=40 keV)~\cite{bib:he10}, and 7.09 MeV ($J^\pi$=4$^+$, $\ell$=2, $\Gamma_p$=80 keV)~\cite{bib:hah96,bib:har02}. The
uncertainties in $E_x$ and $\Gamma_p$ are estimated to be about $\pm$30 keV.}
\begin{ruledtabular}
\begin{tabular}{|c|c|c|}
  Label & $E_x$=7.34 MeV($\Gamma_p$=50 keV) & $E_x$=7.71 MeV($\Gamma_p$=80 keV) \\
  \hline
  fit1  & 2$^+$[$\ell=2$] & 2$^-$[$\ell=3$] \\
  fit2  & 1$^-$[$\ell=3$] &  same as fit1   \\
  fit3  &  same as fit1   & 1$^-$[$\ell=3$] \\
  fit4  &  same as fit1   & 2$^+$[$\ell=0$] \\
  fit5  &  same as fit1   & 3$^-$[$\ell=3$] \\
\end{tabular}
\end{ruledtabular}
\end{table}

\section{Discussion}
Resonant properties of the excited states in $^{18}$Ne are discussed in the following part.

\subsection{6.15-MeV state}
Previously, it was thought the key 1$^-$ resonance in calculating the stellar reaction rate of $^{14}$O($\alpha$,p)$^{17}$F({\em e.g.},
see~\cite{bib:hah96,bib:har02,bib:he09}, {\em etc.}), a recent study~\cite{bib:he10} however, indicated that it was, most probably, a 3$^-$ or 2$^-$ state by
reanalyzing the previous data~\cite{bib:gom01}. This work gives same spin-parity assignments ({\em i.e.}, 3$^-$ or 2$^-$) with a transferred angular
momentum $\ell$=3, but it was an $\ell$=1 used in Ref.~\cite{bib:he10}.
Although a definitive conclusion for its spin-parity assignment can't be given here because of the low statistics both in the present and in the previous {\tt ORNL}
data~\cite{bib:gom01,bib:he10}, this kind of study will shed light on the further experiments. As suggested in Ref.~\cite{bib:he10}, additional high-precision
($^3$He,$n$) or ($p$,$t$) as well as high-statistics ($p$,$p$) elastic-scattering experiments are still strongly required to confirm the spin-parity for this state and
the 6.286-MeV state.

\subsection{7.05-MeV state}
The present experimental data support the peak around 7.05 MeV as a doublet, which is consistent with the previous results~\cite{bib:ner81,bib:hah96}. Now it can be
concluded that one of the doublet, say, the latter 7.09-MeV state is of a 4$^+$ character based on the present and Harss {\em et al.}'s data~\cite{bib:har02},
what's the spin-parity for another state (lower one) then? Two possible assignments (1$^-$ or 2$^+$) proposed by Nero {\em et al.}~\cite{bib:ner81} were attempted in
the present $R$-matrix analysis. It's found that the fitting results of a doublet are better than a singlet. If taking the precise energies from Ref.~\cite{bib:hah96},
the doublet could possibly be a state at 7.05 MeV (1$^-$ or 2$^+$) and another one at 7.12 MeV (4$^+$).
Furthermore, both 7.05- and 7.12-MeV states were observed in the previous thick-target ($\alpha$, $p$) experiment~\cite{bib:not04,bib:not041,bib:kub06}.

\subsection{7.35-MeV state}
Previously, it was observed in the ($^3$He,$n$), ($^{12}$C,$^6$He) reactions~\cite{bib:hah96} and showed (1$^-$, 2$^+$) characters in the measured ($^3$He,$n$) angular
distribution. Hahn {\em et al.}~\cite{bib:hah96} suggested a 1$^-$ for this state based on a very simple mirror argument. Later, following Fortune and Sherr's
arguments~\cite{bib:for00}, Harss {\em et al.}~\cite{bib:har02} speculated it as a 2$^+$ state based on the Coulomb-shift discussion.
Present data support the 2$^+$ assignment rather than the 1$^-$ one~\cite{bib:hah96}.

\subsection{7.71-MeV state}
It was observed in the ($^3$He,$n$), ($^{12}$C,$^6$He) reactions but not the ($p$,$t$) reaction, and behaved an unnatural parity character in the angular
distribution~\cite{bib:hah96}.
In addition, it was also not observed in the ($p$,$\alpha$) reaction~\cite{bib:har02}. Therefore, Harss {\em et al.}~\cite{bib:har02} assigned it as a unnatural 2$^-$
state, and this assignment could reproduce their excitation function data although the structure was not evident in the measured energy region.
The present $R$-matrix calculation gives three possible assignments (3$^-$, 2$^-$, 1$^-$), and hence the previous 2$^-$ assignment is reasonable.
Actually, this 7.71-MeV state was not observed in the ($\alpha$, $p$) reaction~\cite{bib:not04,bib:not041,bib:kub06}, and exhibited an unnatural
character again.

\subsection{other states}
Previously, the 7.60-MeV state was assigned as a natural 1$^-$ state because of its strong population in the $^{17}$F($p$,$\alpha$)$^{14}$O reaction~\cite{bib:har02}.
In this work, a `groove-like' structure was observed at $E_{c.m.}$$\sim$3.6 MeV ({\em i.e.}, $E_{x}$$\sim$7.5 MeV) as seen in Fig.~\ref{fig6}, and it possibly
indicates the existence of a 1$^-$ ($\ell$=1) resonance which may show a `groove' rather than a `bump' structure in the $^{17}$F+$p$ excitation function~\cite{bib:he10}.
But here it can't be fitted reliably due to low statistics.
In addition, we also didn't observe a peak at $E_{c.m.}$$\sim$2.7 MeV, which was thought an inelastic branch of $^{14}$O($\alpha$,p)$^{17}$F reaction proceeding
through a resonance at $E_x\sim$7.1 MeV in $^{18}$Ne which decays to the first excited state in $^{17}$F~\cite{bib:not04,bib:not041,bib:kub06}.
Therefore, the present and the new {\tt ORNL} experimental dat~\cite{bib:bar10} rule out this inelastic-branch interpretation.

\section{Astrophysical implication}
For the stellar reaction of $^{14}$O($\alpha$,p)$^{17}$F, the resonant contribution to the total reaction rate can be calculated by using the isolated and narrow
resonance formula~\cite{bib:wie87,bib:har02,bib:he07,bib:he09},

\begin{widetext}
\begin{eqnarray}
N_A\langle \sigma v \rangle_\mathrm{res} = 1.54\times10^{11}\left(\frac{1}{\mu T_9}\right)^{3/2}\Sigma_i(\omega\gamma)_i \times \exp \left(-\frac{11.605E_r^i}{T_9} \right) \mathrm{[cm^3s^{-1}mol^{-1}]}
\label{eq:three},
\end{eqnarray}
\end{widetext}
where the $E_r^i$ are the center-of-mass energies and the $(\omega\gamma)_i$ are the strengths of the resonances in MeV, the reduced mass $\mu$ in unit of amu.
For this ($\alpha$,$p$) reaction, because of $\Gamma_{\alpha}\ll\Gamma_{p}\approx\Gamma_\mathrm{tot}$~\cite{bib:hah96,bib:har02}, the resonance strength can be calculated
by~\cite{bib:he09},
\begin{eqnarray}
\omega\gamma=\omega\frac{\Gamma_{\alpha}\Gamma_{p}}{\Gamma_\mathrm{tot}}\simeq(2J+1)\Gamma_{\alpha}
\label{eq:four},
\end{eqnarray}
where $J$ is the spin of the resonance. The $\alpha$-particle partial width can be expressed in a form of~\cite{bib:hah96,bib:he09}
\begin{eqnarray}
\Gamma_\alpha \propto  C^2S_\alpha \times P_\ell(E_r)
\label{eq:five}.
\end{eqnarray}
Here $C^2S_{\alpha}$ is the $\alpha$-particle spectroscopic factor~\cite{bib:hah96},
and the Coulomb penetrability factor $P_{\ell}$ is calculated by a {\tt RCWFN} code~\cite{bib:bar74} with same radius of $r_0$=1.25 fm.

\begin{table*}
\caption{\label{table2} Resonant parameters used in the resonant reaction-rate calculations for the stellar $^{14}$O($\alpha$,p)$^{17}$F reaction.}
\begin{ruledtabular}
\begin{tabular}{lcccccccccccc}
 \multicolumn{1}{c}{} & \multicolumn{2}{c}{6.15 MeV} & \multicolumn{2}{c}{6.286 MeV} & \multicolumn{2}{c}{7.05 MeV} & \multicolumn{2}{c}{7.12 MeV} & \multicolumn{2}{c}{7.35 MeV} & \multicolumn{2}{c}{7.60 MeV}\\
 \cline{2-3} \cline{4-5} \cline{6-7} \cline{8-9} \cline{10-11} \cline{12-13}
 & $J^\pi$ & $\omega\gamma$ (eV) & $J^\pi$ & $\omega\gamma$ (eV) & $J^\pi$ & $\omega\gamma$ (eV) & $J^\pi$ & $\omega\gamma$ (eV) & $J^\pi$ & $\omega\gamma$ (eV) & $J^\pi$ & $\omega\gamma$ (eV) \\

\hline
 Hahn {\em et al.}~\cite{bib:hah96}  & 1$^-$ & 6.6  & 3$^-$ & 2.4 & 4$^+$ & 430\footnotemark[1]             &       &                                 & 1$^-$ &4500 &       &      \\
 Harss {\em et al.}~\cite{bib:har02} & 1$^-$ & 9.6  & 3$^-$ & 2.4 & 4$^+$ & 360                             &       &                                 & 2$^+$ & 200 & 1$^-$ & 3000 \\
\hline
 Present \& Ref.~\cite{bib:he10}     &       &      &       &     &       &                                 &       &                                 &       &     &       &\\
 Para1                               & 3$^-$ & 0.36 & 1$^-$ & 40  & 1$^-$ & 5.2$\times 430$\footnotemark[2] & 4$^+$ & 1.5$\times 430$\footnotemark[1] & 2$^+$ & 200 & 1$^-$ & 3000 \\
 Para2                               & 3$^-$ & 0.36 & 1$^-$ & 40  & 2$^+$ & 2.6$\times 430$\footnotemark[2] & 4$^+$ & 1.5$\times 430$\footnotemark[1] & 2$^+$ & 200 & 1$^-$ & 3000 \\
\end{tabular}
\footnotemark[1] An $\alpha$-particle spectroscopic factor $C^2S_\alpha$=0.11 was adopted for the 4$^+$ state~\cite{bib:hah96}. \\
\footnotemark[2] An identical $\alpha$-particle spectroscopic factor $C^2S_\alpha$=0.01 was assumed for the 1$^-$, 2$^+$ assignments~\cite{bib:wie87}.
\end{ruledtabular}
\end{table*}

\begin{figure}[h]
\includegraphics[scale=0.6]{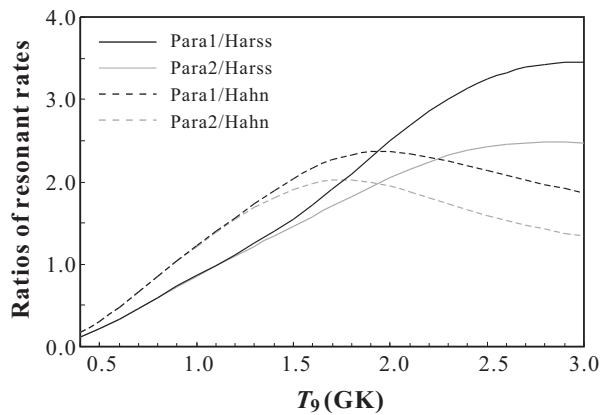}
\caption{\label{fig8} Calculated ratios between the present resonant reaction rates and those previous ones at certain temperature range. The parameters used in the
calculations are listed in Table~\ref{table2}. See text for details.}
\end{figure}

For the broad state at 7.05 MeV, a doublet has been considered in the present calculation, {\em i.e.}, a state at 7.05 MeV (1$^-$ or 2$^+$) and another
at 7.12 MeV (4$^+$) in comparison to a singlet ({\em i.e.}, 7.05 MeV, 4$^+$) used in the previous calculations~\cite{bib:hah96,bib:har02}.
With Equs.~\ref{eq:four}\&\ref{eq:five},
the resonance strength of the latter 7.12-MeV state (4$^+$, $E_r$=2.01 MeV) is about a factor of $R$=$\frac{P_\ell=4(E_r=2.01)}{P_\ell=4(E_r=1.94)}$$\simeq$1.5
larger than the previous value using an energy of 7.05 MeV ($E_r$=1.94 MeV), but the corresponding resonant rate ratio still remains about unity above $\sim$2 GK
because of the exponential term [exp(-11.605$E_r/T_9$)].
The former 7.05-MeV state is possibly of $J^\pi$=1$^-$ or 2$^+$ for which an identical $C^2S_{\alpha}$ factor 0.01 is assumed~\cite{bib:wie87}, and a
$C^2S_{\alpha}$=0.11 for the 7.05-MeV state (4$^+$) is adopted from the previous calculation~\cite{bib:hah96}.
In case of 1$^-$, the strength ratio between the 1$^-$ and 4$^+$ assignments is about
$R$=$\frac{\omega(1^-)}{\omega(4^+)}$$\times$$\frac{C^2S_\alpha(1^-)}{C^2S_\alpha(4^+)} \times \frac{P_\ell=1(E_r=1.94)}{P_\ell=4(E_r=1.94)}$$\simeq$5.2;
in case of 2$^+$, the ratio between the 2$^+$ and 4$^+$ assignments is about
$R$=$\frac{\omega(2^+)}{\omega(4^+)}$$\times$$\frac{C^2S_\alpha(2^+)}{C^2S_\alpha(4^+)} \times \frac{P_\ell=2(E_r=1.94)}{P_\ell=4(E_r=1.94)}$$\simeq$2.6.
Thus, the resonant contribution from the 7.05-MeV `doublet' could be dominated by its lower-energy component and will be enhanced by a factor of about 3.6 or 6.2
(or even larger for a larger $S$ factor, {\em i.e.}, assuming $C^2S_\alpha$$>$0.01) in total.

Totally six resonances have been included in the present calculation, and the parameters are listed in Table~\ref{table2}. Here, the parameters for the 6.15- and
6.286-MeV states are adopted from Ref.~\cite{bib:he10}, and those for the 7.35-MeV and 7.60-MeV states are from Harss {\em et al.}'s work~\cite{bib:har02}.
For the 7.05-MeV state, the parameter sets for the 1$^-$, and 2$^+$ assignments are referred to as `Para1', and `Para2', respectively.
The total resonant reaction-rate ratios between the present and the previous ones are shown in Fig.~\ref{fig8} for resonances listed in Table~\ref{table2}.
It shows that the present resonant rates are considerably larger than the previous ones (either Hahn {\em et al.}'s or Harss {\em et al.}'s) above $\sim$1.5 GK while
smaller than the previous values below $\sim$1 GK. The reasons are mainly due the exchange of spin-parities for the 6.15-MeV and 6.286-MeV state as well as the doublet
assumption for the 7.05-MeV broad state.
The present results probably imply that a considerable enhancement of the resonant reaction rates of $^{14}$O($\alpha$,p)$^{17}$F could be expected at temperature
above $\sim$1.5 GK at which this reaction is thought to be of significant for triggering the $rp$-process in x-ray bursters~\cite{bib:wie87}.
Therefore, determination of spin-parities for the 6.15-, 6.286-MeV states and property of the 7.05-MeV `doublet' becomes very important in calculating the resonant
rates of this crucial reaction.

\section{Summary}
We have investigated the proton resonant properties in $^{18}$Ne through an elastic scattering of $^{17}$F+$p$ by using a radioactive beam of $^{17}$F bombarding a
thick hydrogen target. The resonance
parameters have been determined by the $R$-matrix analysis of the $c.m.$ differential cross sections, and those for the 5.11, 7.35, and 7.71 MeV states are consistent
with the previous results. Interestingly, a $J^\pi$=(3$^-$, 2$^-$) is tentatively assigned to the 6.15-MeV state which was thought the key 1$^-$ state; as for the broad
peak observed at 7.05 MeV, a better fitting $\chi^2$/N value can be achieved by a doublet assignment. Assuming the `doublet' contains a state at 7.05 MeV (1$^-$ or 2$^+$) and
another at 7.12 MeV (4$^+$), the resultant resonant rate of $^{14}$O($\alpha$,p)$^{17}$F (contributing from this doublet) will be enhanced by least a factor of about
3.6 (2$^+$ assumption for the 7.05-MeV state) or 6.2 (1$^-$ assumption for the 7.05-MeV state), and it may imply that this doublet would contribute more to the total rate at relatively low temperature region in which
the 6.15-MeV state was thought dominant before.
In combination with the recent 3$^-$, and 1$^-$ assignments to the 6.15-MeV, and 6.286-MeV states, the present total resonant reaction rates deviate from
those previous ones considerably mainly due the exchange of spin-parities for the 6.15-MeV and 6.286-MeV state as well as the doublet assumption for the 7.05-MeV broad
state.
Therefore, additional high-precision and high-statistics experiments are needed urgently to determine those key resonant parameters, and that
will help us to deeply understand the role of breakout $^{14}$O($\alpha$,p)$^{17}$F reaction occuring in the stellar x-ray bursters.

\vspace{1mm}
We would like to acknowledge the staff of {\tt HIRFL} for operation of the cyclotron and the staff of {\tt RIBLL} for their friendly collaboration.
This work is financially supported by the ``100 Persons Project" (BR091104) and the ``Project of Knowledge Innovation Program" of Chinese Academy of Sciences
(KJCX2-YW-N32), and also supported by the National Natural Science Foundation of China(10975163,11021504), and the Major State Basic Research Development Program of
China (2007CB815000).



\begin{thebibliography}{99}
\bibitem{bib:woo76}
S.E. Woosley and R.E. Taam, Nature 263 (1976) 101.
\bibitem{bib:cha92}
A.E. Champage and M. Wiescher, Annu. Rev. Nucl. Part. Sci. 42 (1992) 39.
\bibitem{bib:wie98}
M. Wiescher, H. Schatz, and A.E. Champagne, Phil. Trans. R. Soc. A 356 (1998) 2105.
\bibitem{bib:taa85}
R.E. Taam, Annu. Rev. Nucl. Part. Sci. 35 (1985) 1.
\bibitem{bib:bar00}
D.W. Bardayan, J.C. Blackmon, C.R. Brune {\it et al.}, Phys. Rev. C 62 (2000) 055804.
\bibitem{bib:sch98}
H. Schatz, A. Aprahamian, J. G$\ddot{o}$rres {\it et al.}, Phys. Rep. 294 (1998) 167.
\bibitem{bib:sch01}
H. Schatz, A. Aprahamian, V. Barnard {\it et al.}, Phys. Rev. Lett. 86 (2001) 3471.
\bibitem{bib:bre09}
M. Breitenfeldt, G. Audi, D. Beck {\it et al.}, Phys. Rev. C 80 (2009) 035805.
\bibitem{bib:wie87}
M. Wiescher, V. Harms, J. G$\ddot{o}$rres {\it et al.}, Astrophys. J. 316 (1987) 162.
\bibitem{bib:fun88}
C. Funck and K. Langanke, Nucl. Phys. A 480 (1988) 1888.
\bibitem{bib:fun89}
C. Funck, B. Grund, and K. Langanke, Z. Phys. A 332 (1989) 109.
\bibitem{bib:hah96}
K.I. Hahn, A. Garc\'{i}a, E.G. Adelberger {\it et al.}, Phys. Rev. C 54 (1996) 1999.
\bibitem{bib:aud03}
G. Audi, A.H. Wapstra and C. Thibault, Nucl. Phys. A 729 (2003) 337.
\bibitem{bib:har99}
B. Harss, J.P. Greene, D. Henderson {\it et al.}, Phys. Rev. Lett. 82 (1999) 3964.
\bibitem{bib:gom01}
J. G\'{o}mez del Campo, A. Galindo-Uribarri, J.R. Beene {\it et al.}, Phys. Rev. Lett. 86 (2001) 43.
\bibitem{bib:har02}
B. Harss, C.L. Jiang, K.E. Rehm {\it et al.}, Phys. Rev. C 65 (2002) 035803.
\bibitem{bib:for00}
H.T. Fortune and R. Sherr, Phys. Rev. Lett. 84 (2000) 1635.
\bibitem{bib:hjj09}
J.J. He, P.J. Woods, T. Davinson {\it et al.}, Phys. Rev. C 80 (2009) 042801(R).
\bibitem{bib:hjj10}
J.J. He, P.J. Woods, T. Davinson {\it et al.}, Nucl. Phys. A 834 (2010) 670(c).
\bibitem{bib:bla03}
J.C. Blackmon, D.W. Bardayan, W. Bradfield-Smith {\it et al.}, Nucl. Phys. A 718 (2003) 127(c).
\bibitem{bib:he10}
J.J. He, H.W. Wang, J. Hu {\it et al.},  (2010), submitted to Phys. Rev. C.
\bibitem{bib:not04}
M. Notani, S. Kubono, T. Teranishi {\it et al.}, Nucl. Phys. A 738 (2004) 411(c).
\bibitem{bib:not041}
M. Notani, S. Kubono, T. Teranishi {\it et al.}, Nucl. Phys. A 746 (2004) 113(c).
\bibitem{bib:kub06}
S. Kubono, T. Teranishi, M. Notani {\it et al.}, Eur. Phys. J. A 27 (2006) 327.
\bibitem{bib:bar10}
D.W. Bardayan, J.C. Blackmon, K.Y. Chae {\it et al.}, Phys. Rev. C 81 (2010) 065802.
\bibitem{bib:art90}
K.P. Artemov, O.P. Belyanin, A.L. Vetoshkin {\it et al.}, Sov. J. Nucl. Phys. 52 (1990) 408.
\bibitem{bib:kub01}
S. Kubono, Nucl. Phys. A 693 (2001) 221.
\bibitem{bib:ter07}
T. Teranishi, S. Kubono, H. Yamaguchi, J.J. He {\it et al.}, Phys. Lett. B 650 (2007) 129.
\bibitem{bib:he07}
J.J. He, S. Kubono, T. Teranishi {\it et al.}, Phys. Rev. C 76 (2007) 055802.
\bibitem{bib:he09}
J.J. He, S. Kubono, T. Teranishi {\it et al.}, Phys. Rev. C 80 (2009) 015801.
\bibitem{bib:xia02}
J.W. Xia, W.L. Zhan, B.W. Wei {\it et al.}, Nucl. Instr. and Meth. Phys. Res. A 488 (2002) 11.
\bibitem{bib:zha10}
W.L. Zhan, H.S. Xu, G.Q. Xiao {\it et al.}, Nucl. Phys. A 834 (2010) 694(c).
\bibitem{bib:sun03}
Z. Sun, W.L. Zhan, Z.Y. Guo {\it et al.}, Nucl. Instr. and Meth. A 503 (2003) 496.
\bibitem{bib:ma10}
P. Ma, J.S. Wang, L.M. Duan {\it et al.}, Atomic Energy Sci. and Tech. (2010), in press.
\bibitem{bib:micron}
Micron Semiconductor Ltd., Lancing, UK. Please see: http://www.micronsemiconductor.co.uk/.
\bibitem{bib:zie85}
J.F. Ziegler {\it et al.}, {\it The Stopping and Range of Ions in Solids}, Pergamon Press, New York, 1985.
\bibitem{bib:lan58}
A.M. Lane and R.G. Thomas, Rev. Mod. Phys. 30 (1958) 257.
\bibitem{bib:des03}
P. Descouvemont, {\it Theoretical Models for Nuclear Astrophysics}, Nova Science Pubishers Inc., New York, 2003.
\bibitem{bib:bru02}
C.R. Brune, Phys. Rev. C 66 (2002) 044611.
\bibitem{bib:alex09}
A.St.J. Murphy, A.M. Laird, C. Angulo {\it et al.}, Phys. Rev. C 79 (2009) 058801.
\bibitem{bib:ner81}
A.V. Nero, E.G. Adelberger and F.S. Dietrich, Phys. Rev. C 24 (1981) 1864.
\bibitem{bib:bar74}
A.R. Barnett {\it et al.}, Comput. Phys. Commun. {\bf 8} (1974) 377.

\end{thebibliography}
\end{document}